\pdfoutput=1
\documentclass[prx,aps,twocolumn,preprintnumbers,amsmath,amssymb,superscriptaddress,showpacs]{revtex4-1}
\usepackage{amsmath}
\usepackage{graphicx}
\usepackage{amssymb}
\usepackage{color}
\usepackage[bookmarks=true,colorlinks=true,linkcolor=blue,urlcolor=blue,citecolor=blue]{hyperref}

\begin{document}

\title{Phase transition of the q-state clock model: duality and tensor renormalization}

\author{Jing Chen}
\affiliation{Institute of Physics, Chinese Academy of Sciences, P.O. Box 603, Beijing 100190, China}
\affiliation{University of Chinese Academy of Sciences, Beijing, 100049, China}

\author{Hai-Jun Liao}
\email{Correspondence author. Email: navyphysics@iphy.ac.cn}
\affiliation{Institute of Physics, Chinese Academy of Sciences, P.O. Box 603, Beijing 100190, China}

\author{Hai-Dong Xie}
\affiliation{Institute of Physics, Chinese Academy of Sciences, P.O. Box 603, Beijing 100190, China}
\affiliation{University of Chinese Academy of Sciences, Beijing, 100049, China}

\author{Xing-Jie Han}
\affiliation{Institute of Physics, Chinese Academy of Sciences, P.O. Box 603, Beijing 100190, China}
\affiliation{University of Chinese Academy of Sciences, Beijing, 100049, China}

\author{Rui-Zhen Huang}
\affiliation{Institute of Physics, Chinese Academy of Sciences, P.O. Box 603, Beijing 100190, China}
\affiliation{University of Chinese Academy of Sciences, Beijing, 100049, China}

\author{Song Cheng}
\affiliation{Institute of Physics, Chinese Academy of Sciences, P.O. Box 603, Beijing 100190, China}
\affiliation{University of Chinese Academy of Sciences, Beijing, 100049, China}

\author{Zhong-Chao Wei}
\affiliation{Institute for Theoretical Physics, University of Cologne, Cologne 50937,Germany}

\author{Zhi-Yuan Xie}
\affiliation{Department of Physics, Renmin University of China, Beijing 100872, China }
\affiliation{Institute of Physics, Chinese Academy of Sciences, P.O. Box 603, Beijing 100190, China}

\author{Tao Xiang}
\email{Correspondence author. Email: txiang@iphy.ac.cn}
\affiliation{Institute of Physics, Chinese Academy of Sciences, P.O. Box 603, Beijing 100190, China}
\affiliation{University of Chinese Academy of Sciences, Beijing, 100049, China}
\affiliation{Collaborative Innovation Center of Quantum Matter, Beijing 100190, China}

\date{\today}

\begin{abstract}
  We investigate the critical behavior and the duality property of the ferromagnetic $q$-state clock model on the square lattice based on the tensor-network formalism. From the entanglement spectra of local tensors defined in the original and dual lattices, we obtain the exact self-dual points for the model with  $q \leq 5 $ and approximate self-dual points for $q \geq 6$. We calculate accurately  the lower and upper critical temperatures for the six-state clock model from the fixed-point tensors determined using the higher-order tensor renormalization group method and compare with other numerical results.
\end{abstract}


\maketitle

Phase transition is one of the most important and challenge problems in condensed matter physics and statistical physics. For a long time, it is believed that all continuous phase transition can be described by Landau symmetry-breaking theory. However, since the early 1970s, a large body of topological phase transitions have been found, which cannot be described by symmetry-breaking. One of the most famous examples is the Kosterlitz-Thouless (KT) transition~\cite{Berezinskii, KT_transition1, KT_transition2} observed in the two-dimensional (2D) XY model. This kind of transitions is driven by topological excitations and does not break any symmetries.

A more interesting model exhibiting KT transition is the $q$-state clock model, which may be considered as a discrete version of the XY model.
This model exhibits a Landau-type second order phase transition between a high temperature paramagnetic phase and a low temperature magnetic long-range ordered phase for $2\leq q\leq 4$  on the square lattice.
When $q \geq 5$, a topological non-trivial KT  phase with quasi long range order emerges between these two phases~\cite{Jose}. Both the transition from the paramagnetic phase to the KT phase and that from the KT phase to the magnetic ordered phase at a lower temperature are of the KT-type. This model has been extensively studied by analytic and numerical methods\cite{Potts, Jose, BA_duality, Cardy, Alcaraz1980, cintia, tobochnik, challa, yamagata, tomita, hwang, brito, baek, kumano, nishino_q6, DMRG_q5}.
However, the phase transition temperatures are difficult to determine due to the topological nature of the KT transition.

Recently, the tensor-network formalism~\cite{TEBD1, TRG, SRG, SRG2, HOTRG, TEFRG, EV_TNR, Loop_TNR} has emerged as a powerful tool to study phases and phase transitions for both quantum and classical systems. Both the generating functionals of quantum systems and the partition functions of classical models can be represented as tensor-network states (or models)~\cite{SRG2}.
By calculating the renormalization flow of the local tensors using the coarse graining tensor renormalization group methods~\cite{TRG, SRG, SRG2, HOTRG, TEFRG, EV_TNR, Loop_TNR}, one can obtain the information on the phases and their critical behaviors.
One can also perform the duality transformation~\cite{Kramers1941} for the tensor network states to establish a relationship between the original and dual systems. In particular, the self-duality can be used to determine the critical points.

In this paper, we investigate the ferromagnetic $q$-state clock model in the tensor network representation by employing the duality transformation and the higher-order tensor renormalization group (HOTRG) method\cite{HOTRG}. We calculate the entanglement spectra of the local tensors defined in the original and dual lattices, and obtain exactly the self-dual points for the model with $q \leq 5 $. For $q \geq 6$, we find an approximate solution for the self-dual points by the duality property of the entanglement entropy of local tensors, which is consistent with the result obtained from the bond-algebraic approach~\cite{BA_duality}. Furthermore, we determine accurately the critical temperatures for the six-state clock model from the fixed-point tensor obtained by the HOTRG iteration.

The ferromagnetic q-state clock model on a square lattice is defined by the Hamiltonian
\begin{equation}
H = -\sum_{\left\langle ij \right\rangle}\cos(\theta_i-\theta_j) ,
\label{eq:Hamitonian}
\end{equation}
where $\theta_i$ is the $i$'th spin angle and takes one of the $q$ discrete values
\begin{equation}
\theta_i = \frac{2\pi}{q}k, \qquad (k = 0,...,q-1).
\end{equation}
This model is also called the vector Potts model. It possesses a discrete $Z_q$ symmetry because the Hamiltonian is invariant if all spin variables $\theta_i$ are changed by $\frac{2\pi}{q}$. If $q = 2$, it reduces to the Ising model. The q = 3 clock model is equivalent to the three-state Potts model. In the limit $q \rightarrow \infty$, it becomes the continuous XY model.

\begin{figure}[t]
\centering
\includegraphics[width=8.0cm]{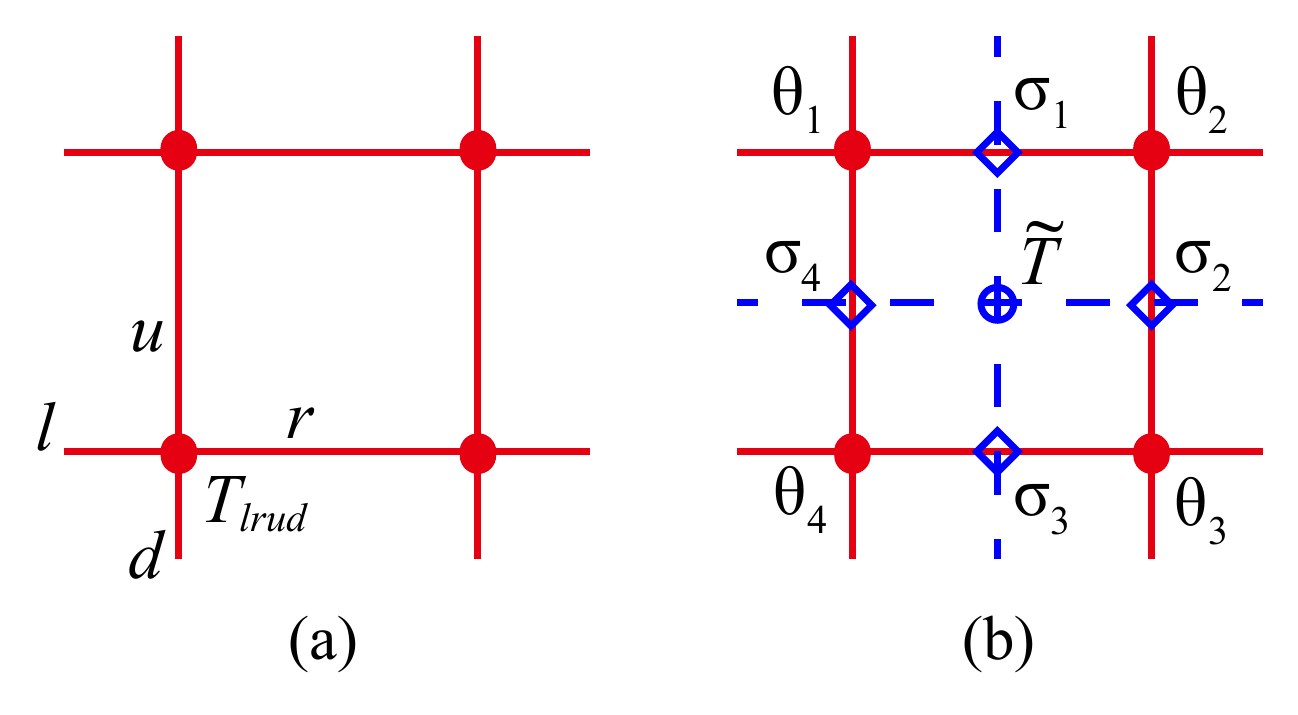}
  \caption{(a) Local tensor $T_{lrud}$ defined in the original square lattice spanned by the red dots.  (b) Local tensor $\tilde{T}_{\sigma_1\sigma_2\sigma_3\sigma_4}$ defined in the dual lattice spanned by the open blue circles. }
\label{Fig:lattice}
\end{figure}

In the original square lattice, the partition function can be expressed as a tensor network model and the local tensor is defined at each lattice site (Fig.~\ref{Fig:lattice}(a))
\begin{equation}
Z=Tre^{-\beta H}=Tr\prod\limits_{i}T_{l_{i},r_{i},u_{i},d_{i}},
\end{equation}
where $\left( l_{i},r_{i},u_{i},d_{i}\right) $, each taking values from $0$ to $q-1$, denote the bond indices linking site $i$ from (left, right, up, down) directions, respectively. The local tensor $T$ is defined by
\begin{equation}
T_{lrud}=\frac{\sqrt{\lambda_{l}\lambda_{r} \lambda_{u}\lambda_{d}}}{q}
\delta_{ \mathrm{mod}(l-r+u-d,q)},
\label{Eq:real_rep}
\end{equation}
where $\lambda_{m}$ ($m=l,r,u,d$) are determined by the singular value decomposition of the Boltzmann weight
\begin{equation}
e^{\beta \cos  (\theta_i-\theta_j)}= \sum_{m} U_{\theta_i,m} \lambda_{m} U_{\theta_j,m}^{\ast },
\end{equation}
and $U_{\theta_i,m} = e^{im\theta_i}$ is an unitary matrix defined at site $i$.  $\lambda_{m}$ is the bond entanglement spectrum between sites $i$ and $j$
\begin{equation}
\lambda_m=\lambda_{-m}=\sum_{\theta_n}e^{-im\theta_n}e^{\beta \mathrm{cos}\theta_n}.
\label{Real_Spectrum}
\end{equation}

\begin{figure}[t]
\centering
\includegraphics[width=8.0cm]{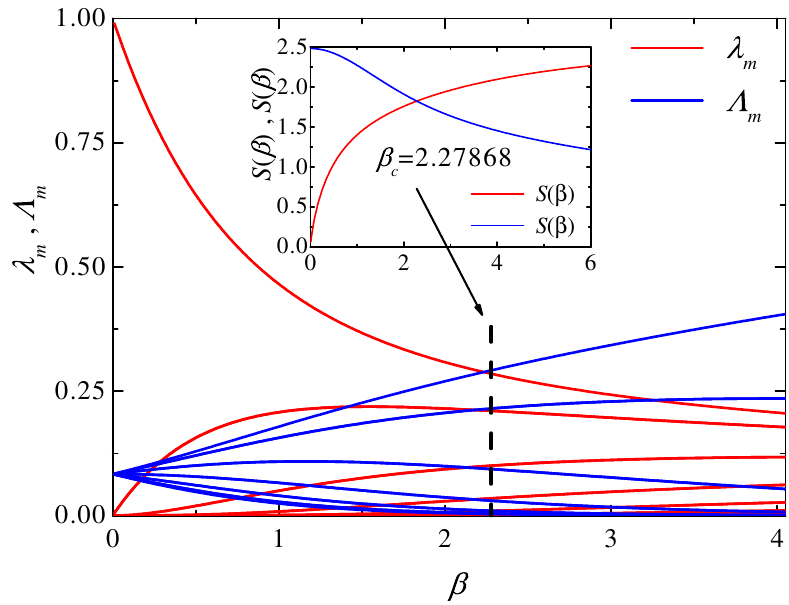}
  \caption{The entanglement spectra for the twelve-state clock model in the original (red) and dual (blue) lattices. Around the approximate self-dual point $\beta_c = 2.27868$ (the dashed black line), determined by the self-dual point of the entanglement entropy (the inset), the bond entanglement spectra $\lambda_{m}$ in the original lattice are approximately degenerate to $\Lambda_{m}$ in the dual lattice. }
\label{Fig:Spectra}
\end{figure}

On the other hand, the partition function can be also written as a product of the Boltzmann weight on all the constitutional small squares
\begin{equation}
Z=Tr\prod\limits_{\left\langle ij\right\rangle \in \square}e^{-\frac{1}{2}\beta H_{ij}},
\end{equation}%
For a small square, we label the spin angles on its four vertexes by $\theta_{1},\theta_{2},\theta_{3}$ and $\theta_{4}$ clockwise as shown in Fig.~\ref{Fig:lattice}(b). On its neighboring squares, we label their vertexes anticlockwise. Then we introduce the following dual spins on each plaquette:
\begin{eqnarray}
\sigma_{1} & = & \theta_{2}-\theta_{1},\nonumber \\
\sigma_{2}& = &\theta_{3}-\theta_{2},\nonumber \\
\sigma_{3}& = &\theta_{4}-\theta_{3},\nonumber \\
\sigma_{4}& = &\theta_{1}-\theta_{4}.\nonumber
\end{eqnarray}
These four dual variables $\sigma_i$ are defined on four bonds of each small square. They are not independent and satisfy the constraint
\begin{equation}
\mathrm{mod}(\sigma_{1}+\sigma_{2}+\sigma_{3}+\sigma_{4},2\pi)=0,
\end{equation}
In the dual lattice, the partition function can be also written as
a tensor-network model
\begin{eqnarray}
  Z & = & \mathrm{Tr} \prod_{\square} \tilde{T}_{\sigma_{1}\sigma_{2}\sigma_{3}\sigma_{4}},
\end{eqnarray}
where $\tilde{T}$ is the local tensor defined at a vertex of the dual lattice (Fig.~\ref{Fig:lattice}(b))
\begin{equation}
  \tilde{T}_{\sigma_{1}\sigma_{2}\sigma_{3}\sigma_{4}} =q\sqrt{\Lambda_{\sigma_1} \Lambda_{\sigma_2} \Lambda_{\sigma_3} \Lambda_{ \sigma_4}} \delta_{\mathrm{mod}(\sum_{i=1}^4 \sigma_i,2\pi)}
\label{Eq:dual_rep}
\end{equation}
with $\Lambda_{\sigma}=e^{\beta \mathrm{cos}\sigma}$ the corresponding bond entanglement spectra in the dual lattice.

\begin{figure}[t]
\centering
\includegraphics[width=8.0cm]{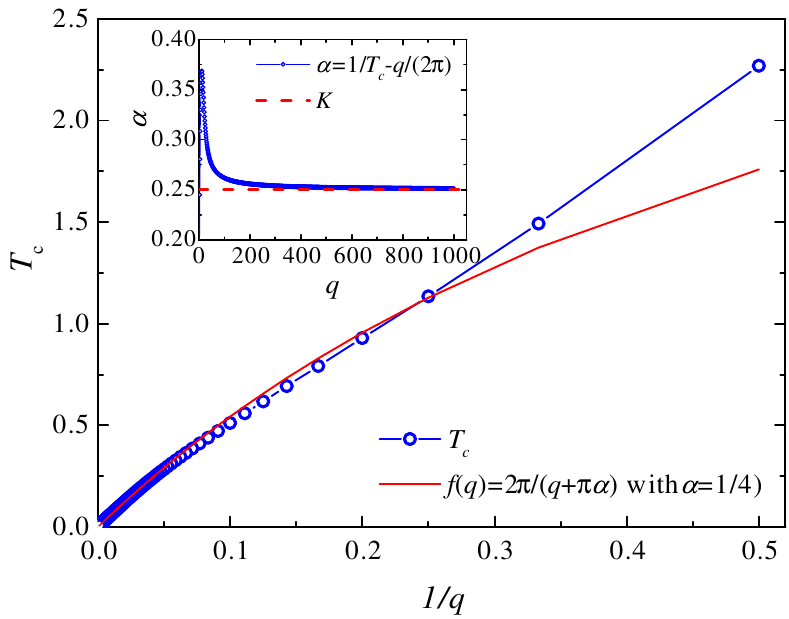}
  \caption{The self-dual temperature, $T_c$ (blue circles), determined from the entanglement entropies of the local tensors defined in the original and dual lattices, as a function of $1/q$.
  The red curve is the asymptotic function, $f(q)=2\pi /(q+ 2\pi \alpha)$, with $\alpha=1/4$. The inset shows the difference $\alpha$ between the self-dual $\beta_c= 1/T_c$ and $q/ 2\pi$.}
\label{Fig:Tc_vs_q}
\end{figure}

The above derivation indicates that the partition function of the clock model is purely determined by the bond entanglement spectra in both the original and dual lattices. Thus the model is dual to each other if the normalized bond entanglement spectra at an inverse temperature $\beta_{1}$ in the original lattice are equal to those at an inverse temperature $\beta_{2}$ in the dual lattice, namely,
\begin{equation}
\frac{\lambda_{m}\left( \beta _{1}\right) }{\lambda_{0}\left( \beta _{1}\right) }=\frac{%
\Lambda_{m}\left( \beta _{2}\right) }{\Lambda_{0}\left( \beta _{2}\right) },\qquad(m=1,\cdots ,q-1).
\label{Eq:dualEqs}
\end{equation}
Setting $\beta_{1}=\beta_{2}=\beta_{c}$, we then obtain the self-dual equation
\begin{equation}
\frac{\lambda_{m}\left( \beta_{c}\right) }{\lambda_{0}\left( \beta_{c}\right) }=\frac{%
\Lambda_{m}\left( \beta_{c}\right) }{\Lambda_{0}\left( \beta_{c}\right) },\qquad(m=1,\cdots ,q-1).
\label{Eq:selfdualEqs}
\end{equation}%
For a system with only one phase transition, the self-dual solution $\beta_{c}$ is just the inverse critical temperature. On the other hand, if the system undergoes more than one transition, $T_c=1/\beta_c$ does not correspond to any of the critical temperatures.

For the $q$-state clock model, it can be shown that Eq.~(\ref{Eq:selfdualEqs}) has an unique solution when $2\leq q\leq 5$. For $2\leq q\leq 4$, the solution is given by
\begin{equation}
\beta_{c}=\left\{
\begin{array}{ccc}
\frac{1}{2}\ln \left( \sqrt{2}+1\right)  &  & q=2, \\
\\
\frac{2}{3}\ln \left( \sqrt{3}+1\right)  &  & q=3, \\
\\
\ln \left( \sqrt{2}+1\right)  &  & q=4,
\end{array}%
\right.
\end{equation}
For $q=5$, the self-dual point $\beta_c$ is determined by the equation
\begin{equation}
\frac{e^{5\beta _{c}/4}}{\cosh \left( \frac{\sqrt{5}}{4}\beta _{c}\right) }= \sqrt{5}+1.
\end{equation}
There is no analytic solution for this equation. By solving this equation numerically, we find that $\beta _{c} \approx 1.076318071604648$.
The above results agree with those obtained by the conventional duality method~\cite{Potts,Alcaraz1980}.
However, the five-state clock model has two transition points. This self-dual solution $1/\beta_c$ does not correspond to any of these two transition temperatures.

\begin{figure}[t]
\centering
\includegraphics[width=8.0cm]{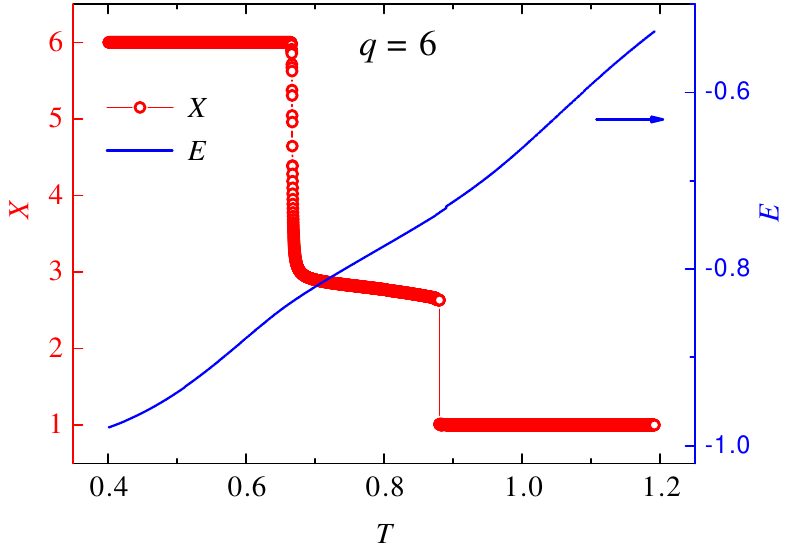}
  \caption{Temperature dependence of parameter $X$ and the internal energy $E$ for the six-state clock model. $X$ exhibits two jumps at the lower and upper critical temperatures, $T_{c1}=0.6658(5)$ and $T_{c2}=0.8804(2)$, respectively. The results are obtained using the HOTRG method with the maximal bond dimension retained for the local tensors $\chi=12$.}
\label{Fig:q6_phase}
\end{figure}

For the clock model with $q\geq 6$, there is no solution for Eq.~(\ref{Eq:selfdualEqs}). However, we can define a bond entanglement entropy via the normalized bond entanglement spectra in the original and dual space, respectively,
\begin{eqnarray}
S(\beta) &=& -\sum_{m}\lambda_{m}\mathrm{ln}(\lambda_{m}), \\
\tilde{S}(\beta) &=& -\sum_{m}\Lambda_{m}\mathrm{ln}(\Lambda_{m}),
\label{eq:entropydual}
\end{eqnarray}
where $\sum_{m}\lambda_{m}=\sum_{m} \Lambda_{m}=1$. At the self-dual point $\beta_c$, apparently $S(\beta_c)=\tilde{S}(\beta_c)$. In case Eq.~(\ref{Eq:selfdualEqs}) does not have a solution, we can still use $S(\beta_c)=\tilde{S}(\beta_c)$ to find an approximate self-dual point, at which the bond entanglement spectra $\lambda_{m}$ in the original lattice are approximately equal to the spectra $\Lambda_{m}$ in the dual lattice. For the $q$-state clock model, we find that this is indeed a good approximation. For example, the bond entanglement spectra of the twelve-state clock model around this self-dual point are approximately degenerate (Fig.~\ref{Fig:Spectra}).
By solving numerically Eq. (\ref{eq:entropydual}), we can find approximately the self-dual temperature $T_c(q) = 1/\beta_c(q)$ for the $q$-state clock model with $q>5$.
Fig.~\ref{Fig:Tc_vs_q} shows the self-dual temperature $T_c$ as a function of $1/q$.
In the large $q$ limit, we find that $\beta_c(q)$ approaches to  $\alpha + q/2\pi$ with $\alpha=1/4$ (see the inset of Fig.~\ref{Fig:Tc_vs_q}). Thus in the limit $q\rightarrow \infty$, $\beta_c(q) \sim q/2\pi$, consistent with the result obtained from the bond-algebraic approach\cite{BA_duality}.

\begin{table}[t]
  \caption{Comparison of the lower and upper critical temperatures, $T_{c1}$ and $T_{c2}$, obtained by different methods for the six-state clock model.}
\label{Table:Tc_List}
\begin{tabular}{lllllllllll}
\hline\hline
&                        ~                         &&&&    $T_{c1}$     &&&&   $T_{c2}$   & \\
\hline
& Tobochnik~\cite{tobochnik}(1982)                 &&&&     0.6         &&&&    1.3       & \\
& Challa and Landau~\cite{challa}(1986)            &&&&     0.68(2)     &&&&    0.92(1)   & \\
& Yamagata and Ono~\cite{yamagata}(1991)           &&&&     0.68        &&&&    0.90      & \\
& Tomita and Okabe~\cite{tomita}(2002)             &&&&     0.7014(11)  &&&&    0.9008(6) & \\
& Hwang~\cite{hwang}(2009)                         &&&&     0.632(2)    &&&&    0.997(2)  & \\
& Brito {\it et al.}~\cite{brito}(2010)            &&&&     0.68(1)     &&&&    0.90(1)   & \\
& Baek {\it et al.}~\cite{baek}(2013)              &&&&       -         &&&&    0.9020(5) & \\
& Kumano {\it et al.}~\cite{kumano}(2013)          &&&&     0.700(4)    &&&&    0.904(5)  & \\
& Kr\v cm\'ar {\it et al.}~\cite{nishino_q6}(2016) &&&&     0.70        &&&&    0.88      & \\
\hline
& this work                                        &&&&     0.6658(5)   &&&&    0.8804(2) & \\
\hline\hline
\end{tabular}
\end{table}

The clock model of $q\ge 5$ undergoes two phase transition from high temperature to low temperature\cite{BA_duality}. The self-dual condition only provides one constraint for the phase boundary, thus cannot be used to determine the two critical points. As there is no local order parameter to describe the intermediate KT phase, it is also difficult to determine the critical points by calculating magnetization, specific heat, or other thermodynamic quantities. To resolve this difficulty, we utilize the HOTRG~\cite{HOTRG} to explore the scaling behavior of the local tensors under the change of length scales. From the critical behavior of the fixed point tensor, we determine very accurately the critical points. The HOTRG handles directly an infinite lattice. The results obtained with this method are not affected by the finite lattice size effect.

The HOTRG is an efficient and flexible tensor renormalization group method. It works in both two and higher dimensions. It has provided accurate results for the two-dimensional XY model~\cite{Yu2014}, three-dimensional Ising~\cite{HOTRG} and Potts models~\cite{WangShun}. At each step of the HOTRG iteration, an optimized isometric matrix is determined by the higher-order singular value decomposition to truncate the bond dimension of local tensors. Eventually, each local tensor will flow to a corresponding fixed-point tensor. One can determine the phase boundary from this fixed-point tensor by calculating the following gauge invariant quantity\cite{TEFRG},
\begin{eqnarray}
X & =& \frac{\left(\sum\limits_{ru}T_{rruu}\right)^{2}} {\sum\limits_{lrud}T_{rlud}T_{lrdu}},
\end{eqnarray}
where $X$ is an effective measure of the degeneracy of the phase.

For the $q$-state clock model, we find that $X$ equals $q$ in the low temperature symmetry-breaking phase, and $1$ in the high temperature disordered phase. In the intermediate KT critical phase, $X$ takes values between $q$ and $1$. $X$ exhibits two jumps at the two phase boundaries, which can be used to determine the critical points.
Fig.~\ref{Fig:q6_phase} shows, as an example, the temperature $T$ dependence of $X$ as well as the internal energy obtained by the HOTRG calculation with the number of states retained at each bond $\chi=12$ for the $q=6$ clock model. From the critical behavior of $X$, we find the lower and upper critical temperatures to be $T_{c1}=0.6658(5)$ and $T_{c2}=0.8804(2)$, respectively. The slops of the internal energy are slightly changed around the jumps. These results are consistent with other numerical calculations~\cite{tobochnik, challa, yamagata, tomita, hwang, brito,baek, kumano,nishino_q6,DMRG_q5}. A comparison for the lower and upper critical temperatures is shown in Table.~\ref{Table:Tc_List}.

In summary, we propose a simple tensor-network scheme to analyze the duality properties of the ferromagnetic $q$-state clock model, and calculate the critical temperatures using the HOTRG method. From the entanglement spectra of the local tensors in the original and dual lattices, we show that there is a self-dual point for the system with $q \leq 5 $. The self-dual inverse temperatures such obtained for the $q\le 5$ models agree with the exact results previously known. For the model with $q \geq 6$, there is no self-dual point in the entanglement spectra. But we find that the entanglement spectra are approximately self-dual at the self-dual point of the entanglement entropy. We use this self-dual point of the entanglement entropy as an approximate self-dual point for the partition function. In the large $q$ limit, this approximate self-dual point $\beta_c$ scales as $q/2\pi + 1/4$, consistent with the result obtained from the bond-algebraic approach~\cite{BA_duality}. We calculate the critical temperatures for the six-state clock model from the fixed-point tensor obtained by the HOTRG method. Our results are consistent with other numerical calculations.

This work was supported by the National Natural Science Foundation of China Grant No.~11474331.

\end{document}